# High-pressure single-crystal X-ray diffraction study of ErVO$_4$

*Josu Sánchez-Martín,[†] Gastón Garbarino,[‡] Samuel Gallego-Parra,[‡] Alfonso Muñoz,[§] Sushree Sarita Sahoo,[⊥] Kanchana Venkatakrishnan,[⊥] Ganapathy Vaitheeswaran,[⊥] and Daniel Errandonea[\*,†]*

[†]Departamento de Física Aplicada-ICMUV, MALTA Consolider Team, Universidad de Valencia, Dr. Moliner 50, Burjassot, 46100 Valencia, Spain

[‡] European Synchrotron Radiation Facility, Grenoble 38043, France

[§]Departamento de Física, MALTA-Consolider Team, Instituto de Materiales y Nanotecnología, Universidad de La Laguna, San Cristóbal de La Laguna, E-38200 Tenerife, Spain

[⊥] School of Physics, University of Hyderabad, Prof. C. R. Rao Road, Gachibowli, Hyderabad, Telangana 500046, India

*Corresponding author email: daniel.errandonea@uv.es

ABSTRACT: We present an investigation into the crystal structure of ErVO$_4$ under variable pressure conditions. The high-pressure single crystal X-ray diffraction experiments performed employing helium as the pressure medium facilitated structure refinements up to 24.1(2) GPa. The transition from zircon to scheelite was observed at a pressure of 7.9(1) GPa. In contrast to previous reports, we did not detect any sign of phase coexistence. We also did not observe the second phase transitions previously predicted by density-functional theory to occur below 20 GPa. The determination of the pressure dependence of unit-cell parameters and volume yields precise values for linear compressibility of each axis and the pressure-volume equation of state for both the zircon and scheelite phases. Additional information on the mechanical properties of ErVO$_4$, obtained from density-functional theory calculations, is also reported.

## 1. Introduction

The orthovanadates of rare-earth elements, characterized by the formula RVO$_4$ (where R denotes a rare-earth element), constitute a recognized class of ternary oxides. They are a group of materials exhibiting remarkable optical, chemical, and mechanical characteristics. As a result, they have attracted considerable attention in recent decades due to their remarkable technological applications as it can be seen in multiple articles available in the literature [1-4]. The majority of RVO$_4$ compounds display a tetragonal structure that is isomorphic to the crystal arrangement of zircon, classified under the space group *I*4$_1$/*amd* [5] and it is represented in **Figure** 1. RVO$_4$ compounds have been the object of numerous high-pressure studies [6]. It has been found that pressures below 10 GPa are enough to induce phase transitions [6]. These transitions have a significant impact on the physical properties of RVO$_4$ compounds, resulting in notable alterations, including an abrupt decrease of the band-gap energy. [7]. Compounds that include larger trivalent cations, such as cerium (Ce) and praseodymium (Pr), undergo a phase transition to a monoclinic structure referred to as the monazite-type, characterized by the space group *P*2$_1$/*n* [8]. In contrast, compounds featuring smaller cationic radii, specifically the lanthanides ranging from lutetium to neodymium, undergo a transformation into a tetragonal scheelite-type structure characterized by by space group *I*4$_1$/*a* [8]. The conversion of zircon to scheelite in rare-earth orthovanadates is a non-reversible process, resembling the phase transition experienced by the mineral zircon (ZrSiO$_4$). This mineral is a common accessory found in various types of sedimentary, igneous, and metamorphic rocks [9]. The zircon-scheelite transition has been also reported in orthophosphates [10]. However, the transition in vanadates occurs below 10 GPa and in ZrSiO$_4$ and in zircon-type phosphates the phase transition takes place beyond 20 GPa [11]. This makes zircon-type vanadates better candidates to systematically study the zircon-scheelite transition under quasi-hydrostatic conditions.

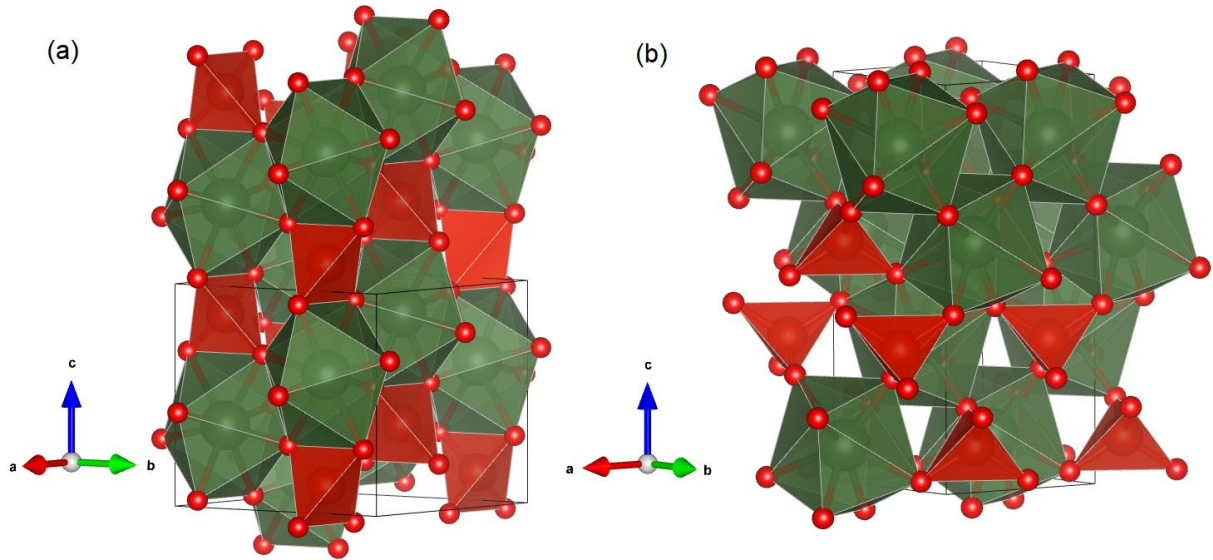

**Figure 1**. The crystal structures of $ErVO_4$ are represented in the two forms: the zircon-type (a) and the scheelite-type (b). In these representations, the $ErO_8$ polyhedral units are depicted in green, while the $VO_4$ tetrahedra are illustrated in red. The small red circles indicate the presence of oxygen atoms. The boundaries of the unit cell are outlined with black lines.

Previous powder x-ray diffraction (XRD) studies have shown that the zircon-scheelite transition is sluggish. The low-pressure zircon-type phase and the high-pressure scheelite-type phase coexist across a wide range of pressures. [12]. For instance, according to powder XRD, in $GdVO_4$, the transition begins at a pressure of 5 GPa; however, it does not fully complete until reaching 23 GPa. [12]. Based on these results a martensitic mechanism was proposed for the phase transition [12]. However, single-crystal XRD studies in related oxides have found that transitions which are characterized by extensive areas of phase coexistence in powder X-ray diffraction experiments do not necessarily show such phase coexistence when experiments are carried out using single crystals [13, 14]. This indicates that the extensive phase coexistence noted in powder experiments involving zircon-type oxides may not stem from an intrinsic source but rather is likely a result of stress interactions between grains. [13,14]. Another issue potentially affecting the zircon-scheelite transition is the influence of non-hydrostatic stress due to the use in experiments of a pressure-transmitting medium (PTM) different than helium. Such stresses could even modify the structural sequence, as it occurs in compounds like $NdVO_4$ [15,16]. Single-crystal XRD studies using helium as PTM have never been performed yet in zircon-type vanadates and it is timely to perform them. The findings will aid not only in enhancing the understanding of the zircon-scheelite transition but also in verifying or refuting the recent suggestion of an intermediate bridge phase existing between zircon and scheelite [17].

Here we will report a high-pressure single crystal study of $ErVO_4$ conducted under conditions that approximate hydrostatic equilibrium up to 24.1(2) GPa using helium as PTM, which provides the best possible experimental conditions. The experiments were supplemented by density-functional theory (DFT) calculations that analyse the changes in the crystal structure due to compression, as well as the elastic constants of the two crystal structures of interest for this work. This compound was previously studied by powder XRD utilizing a mixture of methanol and ethanol in a 4:1 ratio as the PTM [18]. In that research, it was determined that the zircon-scheelite transition begins at a pressure of 8.2 GPa, but a pure scheelite-type phase was not found up to 15.5 GPa. In the same study, the crystal structure of the scheelite structure was solved at 9 GPa by single-crystal XRD in experiments that used neon as PTM. A second high-pressure study in $ErVO_4$ has been reported [19]. In this study the compound was characterized by luminescence measurements performed using argon as PTM. Three phase transitions were observed. The first one occurs at 7.9 GPa, the second one at 20 GPa, and the third one at 31 GPa [19]. The second transition was also predicted by DFT calculations, which predicted a scheelite-fergusonite transition at 18 GPa [8]. In our quasi-hydrostatic single-crystal XRD we have found that the zircon-scheelite transition is not sluggish. We have also excluded the possibility of an intermediate phase existing between zircon and scheelite. We will additionally present the dependence of unit-cell parameters and volume on pressure, and we will compare our findings with prior experimental data as well as current DFT results. The elastic constants and moduli of the two phases will be presented too.

## 2. Materials and methods

The single crystals of $ErVO_4$ were synthesized utilizing the flux method as described by Ruiz-Fuertes *et al*. [18]. The process entails dissolving $Er_2O_3$ in molten $Pb_2V_2O_7$ at 1170 °C, followed by spontaneous nucleation and crystal growth of $ErVO_4$, which occurs through the gradual cooling of the solution. The growth crystals have the zircon-type structure described by space group *I4$_1$/amd* with unit-cell parameters a = 7.0951(7) Å and c = 6.2705(6) Å as confirmed by our single-crystal XRD experiments.

High-pressure single-crystal XRD experiments were performed on beamline ID15B of the ESRF [20]. The data was collected using an Eiger2 X CdTe 9M (DECTRIS) detector with an X-ray wavelength of 0.41 Å and a beam size of 1 × 1 μm. Sample-to-detector distance of 180.79 mm was calibrated using the single-crystal vanadinite standard [21]. Two samples (180 × 80 × 20 μm) were loaded into a membrane-type diamond anvil cell with an aperture angle of 60°. A 200 μm stainless steel gasket was indented to a thickness of 90 μm and a 300 μm diameter chamber was drilled in the center. The diamond culet size was 600 μm. A ruby sphere was placed in the chamber as pressure gauge [22]. Helium was used as the PTM. It was loaded into the diamond-anvil cell using a gas loading system manufactured by Sanchez Technologies, model GLS1500, which is a available at the sample environment service-HP of the ESRF. The collected XRD images were processed with CrysAlis[Pro] (available from Rigaku Americas Corporation) in the following manner. First, the reflections belonging to the same structure were identified, and the rest were rejected. Then, the chosen reflections were reduced using the space group restrictions of the most likely structure (offered by the software using the lower $R_{int}$ parameters). The structural analysis was subsequently conducted utilizing the OLEX2 program [23-25], with the chosen space group and the indexed reflection list being imported. Given these restrictions and the stoichiometry of the sample, the position of atoms was assigned in the regions of highest electronic density, following the demonstration of stability in the atom location.

Ab initio total-energy calculations were conducted using the density-functional theory (DFT), specifically employing the Vienna Ab initio Simulation Package (VASP) [26]. The projector augmented wave (PAW) pseudopotential was utilized, with a plane-wave kinetic cutoff set at 520 eV. For the oxygen atom, a $2s^22p^4$ electronic configuration was considered, while vanadium was represented with a $3p^63d^44s^1$ configuration [27]. In the case of erbium, the Er_3 pseudopotential was applied, with the f electrons treated as frozen in the core [28]. The exchange-correlation energy was described using the generalized gradient approximation (GGA) according to the AM05 Armiento and Mattsson prescription [29.30]. The Perdew-Burke-Ernzerhof (PBE) [31] and PBE for solids [32] were also tested, but we found that AM05 describes better the crystal structure at 0 GPa and therefore we used this approximation for the rest of calculations. The Brillouin zones (BZs) were sampled using dense grids of special k points [33], specifically 6×6×6 and 8×8×8 meshes for zircon, scheelite, and monazite, respectively. This approach achieved a convergence of 1–2 meV per formula unit in total energy and ensured precise calculations of atomic forces. The unit cell parameters and atomic positions underwent full optimization, adhering to criteria that required atomic forces to be less than 0.005 eV/Å and the stress tensor deviations from the diagonal hydrostatic form to be below 0.1 GPa. The simulations were conducted at zero temperature. The differences in enthalpy facilitate the assessment of the relative stability among the various phases of each compound. Lattice-dynamic calculations were executed at the zone centre (Γ point) of the Brillouin zone using the direct force-constant method [34] to investigate dynamical stability of the zircon-type structure.

## 3. Results and discussion

### 3.1 Zircon-scheelite transition

The supplementary material (SM) contains information regarding the data collections, refinement outcomes, and structural data at various pressures, which can be found in Tables S1 and S2. The atomic positions of the structures are detailed in Tables S3 and S4 of the SM. Additionally, CIF files encompassing all structural information have been submitted to the Cambridge Crystallographic Data Centre, with the corresponding CCDC deposition numbers listed in Tables 1, S1, and S2. Two samples were subjected to simultaneous measurement from ambient pressure to 24.1(2) GPa, and the sample that exhibited a single-crystal signal following the phase transition was selected for subsequent analysis. The unit-cell parameters derived from our single-crystal X-ray diffraction experiments across all pressure conditions we have measured are summarized in Table S5 of the SM.

In **Figure** 2, a section of the reciprocal space is presented, specifically from the ($a^*1c^*$) reciprocal layer at pressures of 7.5(1) and 7.9(1) GPa. Our findings indicate that all experiments conducted pressures below 7.5(1) GPa can be clearly ascribed to the zircon-type structure previously recognized through powder X-ray diffraction (XRD) [35] and neutron diffraction [36] at ambient pressure and temperature conditions. The structural framework of zircon-type ErVO4, illustrated in **Figure** 1(a), is characterized by chains composed of alternating edge-sharing VO4 tetrahedra and EuO8 triangular dodecahedra, which run parallel to the c-axis. These chains are laterally interconnected through edge-sharing interactions between the dodecahedra.

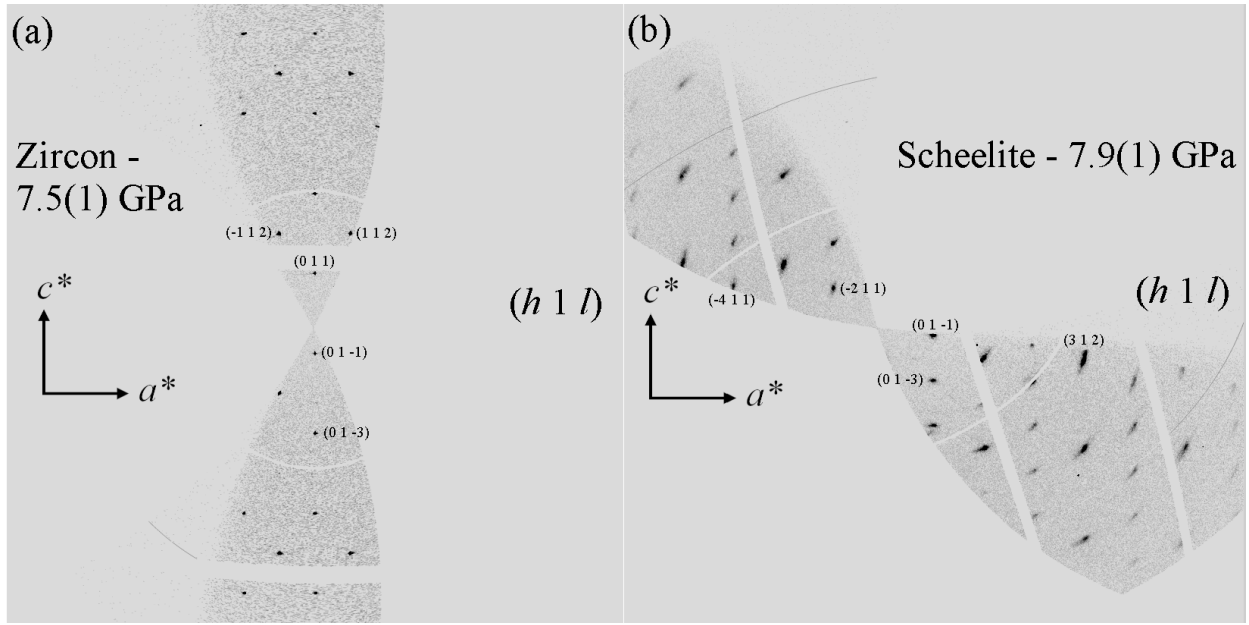

**Figure 2.** Images depicting a portion of the reciprocal space within the ($h1l$) plane for zircon-type ErVO4 (a) and scheelite-type ErVO4 (b) at pressures of 7.5(1) and 7.9(1) GPa, respectively. The alterations observed in the patterns suggest the presence of a phase transition.

At 7.9(1) GPa there are more reflections than at 7.5(1) GPa indicating a decrease in the symmetry of the crystal structure, which is accompanied by a change in the space group from $I4_1/amd$ to $I4_1/a$. We found that at 7.9(1) GPa there are reflections which violates the condition $2h+l = 4n$ (when hhl) that should be satisfied in space group $I4_1/amd$. In contrast all reflections satisfy the reflection conditions for space group $I4_1/a$. The change in the space group reflects a structural transformation that occurs in the material during the phase transition. The analysis of the image collected at 7.9(1) confirmed it can be assigned to the scheelite-type structure in agreement with Ref. 18. From 7.9(1) to 24.1(2) GPa the crystal structure has been identified as isomorphous with scheelite. The crystal structure of the HP phase at 7.9 GPa is reported in **Table** 1. The transition pressure of 7.9(1) GPa agrees with the transition pressure from previous experiments [18,19]. DFT calculations indicated that the enthalpy of the scheelite phase decreases below that of the zircon phase at a pressure of 4.5 GPa, as illustrated in **Figure** 3. According to thermodynamic arguments, this supports a transition pressure of 4.5 GPa. The variation in the transition pressure may be attributed to the calculations being performed at 0 K or to the constraints of DFT in accurately representing the exchange-correlation energy in systems containing f-electron atoms, such as the lanthanides [37]. A further potential explanation for the variation in transition pressure may be associated with the existence of a kinetic barrier that blocks the transition until a pressure value is reached at which the zircon structure becomes dynamically unstable [8]. This argument is consistent with our phonon results shown in the inset of **Figure** 3. Our calculations indicate that the silent $B_{1u}$ mode of the zircon structure transitions to an imaginary state at 8.1 GPa, which is consistent with the experimentally observed transition pressure of 7.9(1) GPa. The softening of the $B_{1u}$ modes initiates a dynamic instability that mitigates the influence of kinetic barriers during the zircon-to-scheelite phase transition, thereby promoting the realization of the observed transition.

**Table 1:** The crystal structure details of the tetragonal scheelite $ErVO_4$ (space group $I4_1/a$) at a pressure of 7.9(1) GPa are provided. Additional information can be found in Table S2 of the SM.

| a,b/Å | c/Å | V/Å³ | Z | $R_{int}$ | GooF | $R_1$ | $wR_2$ | CCDC # |
|---|---|---|---|---|---|---|---|---|
| 4.9470(10) | 10.960(2) | 268.22(12) | 4 | 0.0568 | 0.986 | 0.0731 | 0.1919 | 2393949 |
| Er | x | 0 | V | x | 0 | O | x | 0.156(3) |
|  | y | 0.25 |  | y | 0.25 |  | y | 0.509(3) |
|  | z | 0.625 |  | z | 0.125 |  | z | 0.2078(16) |

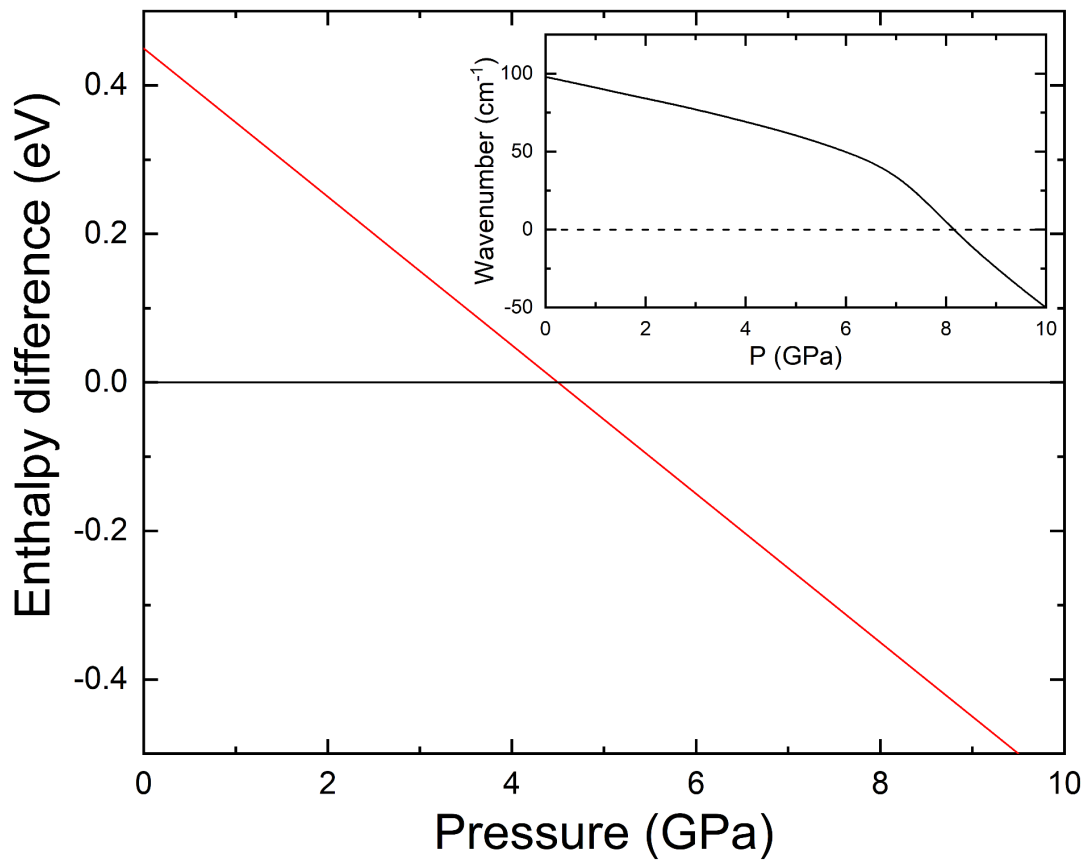

**Figure 3.** Difference between the enthalpy of the scheelite (red) and zircon (black) phase versus pressure. Scheelite becomes the lowest enthalpy structure at 4.6 GPa. The inset shows the wavenumber of the silent $B_{1u}$ mode, which becomes imaginary at 8.1 GPa.

The crystal structure of the phase emerging under high-pressure conditions and characterized as scheelite agrees with that previously reported by Ruiz-Fuertes *et al.* [18] and with that measured in a sample synthesized at 3.5 GPa and 1000-1500 K using a Belt-type apparatus [38]. As shown in **Figure** 1(b) the structure contains Eu atoms coordinated by eight $VO_4$ tetrahedra, sharing an oxygen atom with each tetrahedron. The $EuO_8$ polyhedra and $VO_4$ tetrahedra are connected via common vertices. Each $EuO_8$ polyhedron is edge-sharing with four nearest $EuO_8$ polyhedra.

An interesting characteristic feature of the zircon-scheelite transition we detected in our experiments is that it occurs very fast and that no phase coexistence is observed. In contrast to the results of previous powder XRD and photoluminescence experiments [18, 19], which were conducted in powder samples using pressure-transmitting media that are less hydrostatic than He (the PTM used in our study), a significant pressure range of phase coexistence was observed. This suggests that the phase coexistence noted in previous experiments is not a fundamental characteristic of the zircon-scheelite transition but is likely a result of either the stress between grains or non-hydrostatic conditions (or the combination of both facts), thereby emphasizing the significance of conducting high-pressure single-crystal XRD experiments.

We also found that the observed transition is irreversible, and the unit-cell volume of the scheelite-type structure is 9.9% smaller that of the zircon-type structure. Additionally, our results exclude that between zircon and scheelite it could exist an intermediate phase, which was proposed as a possible mechanism to overcome kinetic barriers [17]. Regarding the transition mechanism our results are supportive of the mechanism proposed by Marqueño *et al.* [8]. In **Figure** 4 we present a projection of the zircon and scheelite structure including selected planes showing that both structures form hexagonal layers. The atomic configuration in both structures exhibits similarities; however, there is a distinct difference in the orientation of the $VO_4$ tetrahedra, which experience both rotation and tilting. The HP scheelite-type structure can be obtained from the zircon-type structure by a shift of all the cations in the layer in the same direction and a tilting of the $VO_4$ tetrahedra. The atomic reorganization favours the sudden decrease of the volume at the phase transition. To conclude this section of the discussion, we wish to emphasize that we found no evidence of a second phase transition in $ErVO_4$ up to 24.0(2) GPa. Such transition was previously detected by photoluminescence at 20 GPa in experiments performed using argon (Ar) as pressure medium [19]. Ar solidifies in the fcc phase at 1.3 GPa being solid Ar very compressible, with a bulk modulus of 6.5(5) GPa [39]. It has been broadly used as quasi-hydrostatic pressure medium; however, the initial indications of pressure gradients are observed at 2 GPa [40]. These gradients progressively rise with increasing pressure, attaining a value of 0.1 GPa at 10 GPa and appear to escalate more swiftly beyond 20 GPa, which is the pressure where the second transition was previously reported [19]. Thus, it is quite likely that the second transition was previously observed in the previous study [19] due to the influence in experiments of non-hydrostatic conditions, which are known to reduce the pressure of phase transitions in zircon-type vanadates [41].

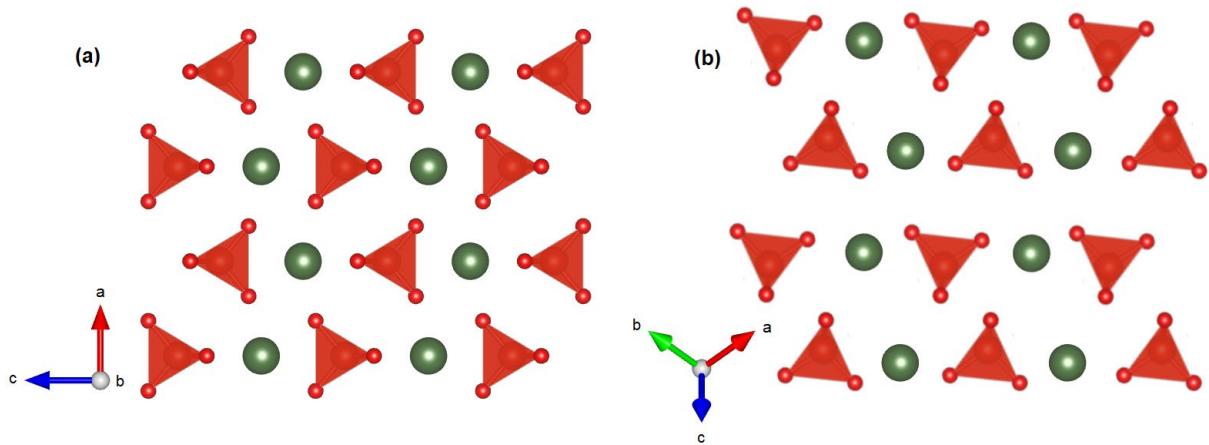

**Figure 4.** Schematic representation of projections of the zircon-type (a) and scheelite-type (b) structures of ErVO$_4$ illustrating their relationship. Scheelite can be obtained from zircon by a shift of all the cations in layer in the same direction and a tilting of VO$_4$ tetrahedra.

3.2 *Room-temperature equation of state*

From the analysis of experiments performed on sample compression and decompression, we established the relationship between pressure and the unit-cell parameters for the two phases of ErVO$_4$, as demonstrated in **Figure** 5. All experimental data used is gathered in Table S5 of the SM. This figure also includes results from DFT calculations as well as data from prior experiments for comparative analysis. For the zircon phase, at ambient conditions, the DFT calculations underestimate by 1% the unit-cell parameter $c$ and agree with experiments within experimental uncertainties regarding the unit-cell parameter $a$. The pressure dependence of unit-cell parameters obtained from current experiments has been contrasted with prior experiments [18] alongside DFT calculations yielding remarkably consistent results. The current experiments reveal only a minor difference when compared to earlier studies [18] concerning the compressibility of the $a$-axis in the zircon phase, with the previous research indicating a slightly lower value than that observed in the present findings (see **Figure** 5). The pressure dependence of unit-cell parameters is linear. From our results we obtained the linear compressibility of each axis of zircon-type ErVO$_4$, $\kappa_a = 2.5(1) \times 10^{-3}$ GPa$^{-1}$ and $\kappa_c = 1.6(1) \times 10^{-3}$ GPa$^{-1}$, and scheelite-type ErVO$_4$, $\kappa_a = 1.5(1) \times 10^{-3}$ GPa$^{-1}$ and $\kappa_c = 2.0(1) \times 10^{-3}$ GPa$^{-1}$. The reported values support an anisotropic compressibility, which is a characteristic behaviour observed in orthovanadates [6]. The findings indicate that in both phases, the longest axes display the greatest compressibility. This phenomenon arises from the fact that in both structures, the VO$_4$ tetrahedra display considerably lower compressibility compared to the ErO$_8$ dodecahedra, which predominantly affect compressibility along the $a$-axis in zircon and the $c$-axis in scheelite [18]. From XRD data, our research established the relationship between pressure and the volume of both polyhedral units. Utilizing a second-order Birch-Murnaghan equation of state [42], we derived a bulk modulus of 135(3) GPa for the ErO$_8$ dodecahedra and a bulk modulus of 250(3) GPa for the VO$_4$ tetrahedra within the zircon phase. On the other hand, we got a bulk modulus of 152(3) GPa for the ErO$_8$ dodecahedra and a value of 255(3) GPa for the bulk modulus of the VO$_4$ tetrahedra in the scheelite phase. The findings corroborate the earlier conclusions presented by Ruiz-Fuertes *et al.* [18] concerning the relationship between polyhedral compressibility and the anisotropic compressibility observed in the zircon and scheelite structures of ErVO$_4$.

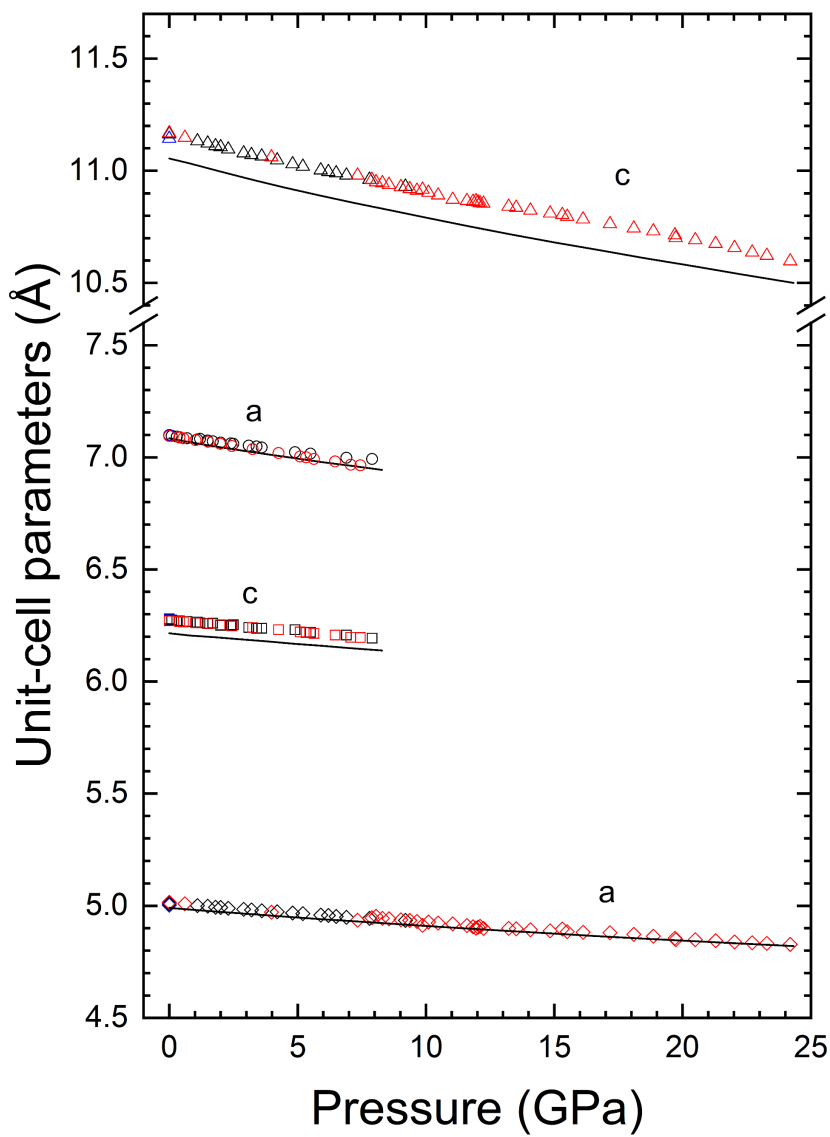

**Figure 5.** The dependence of unit-cell parameters on pressure is illustrated. The circles and squares (as well as triangles and diamonds) represent findings for the zircon (scheelite) phase. The red symbols denote data from the current study, while the black symbols are taken from Reference 18. The blue symbols indicate results obtained at ambient pressure from References 35 and 38. The lines correspond to the outcomes of our DFT calculations. The errors associated with these measurements are less than the dimensions of the symbols used.

**Figure 6** illustrates the relationship between pressure and the unit-cell volume. Our analysis indicates that this relationship can be accurately represented by a second-order Birch-Murnaghan equation of state [42]. Based on the data presented in **Figure 6**, the experimental values for the zircon-type phase at zero pressure are $V_0 = 316.0(1)$ Å$^3$ for the unit-cell volume and $B_0 = 134(2)$ GPa for the bulk modulus. The calculated bulk modulus is consistent with the results obtained from our theoretical results, $V_0 = 313.1$ Å$^3$ and $B_0 = 135.1$ GPa. This agreement suggests that in the previous XRD study [18], the reported bulk modulus, $B_0 = 158(13)$ GPa, was slightly overestimated. For the HP scheelite-type phase we determined from experiments $V_0 = 281.2(4)$ Å$^3$ and $B_0 = 146(3)$ GPa. The bulk modulus agrees within uncertainties with the results from present calculations $V_0 = 276.1$ Å$^3$ and $B_0 = 150.3$ GPa and previous experiments $B_0 = 158(17)$ GPa [18]. The augment in the bulk modulus during the phase transition from 134(2) GPa to 146(3) GPa aligns with the sudden reduction in volume (and corresponding increase in density) observed at the transition.

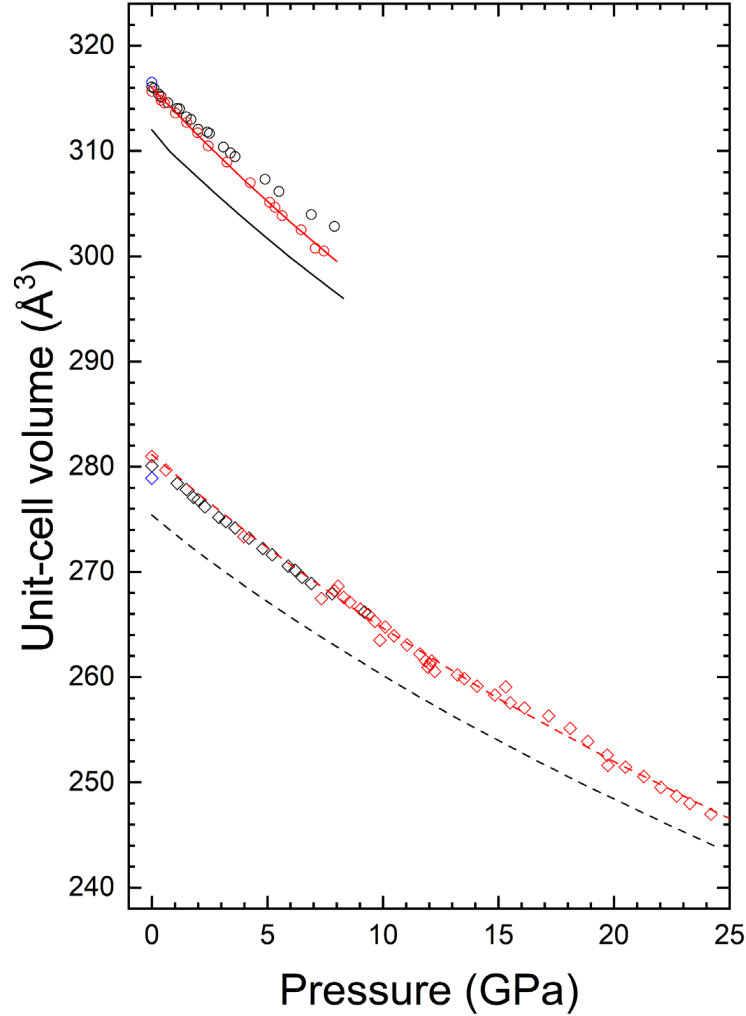

**Figure 6.** The relationship between pressure and the unit-cell volume is illustrated. The circles (and diamonds) represent findings for the zircon (and scheelite) phases. The red symbols denote data from the current investigation, while the black symbols are reproduced from Reference 18. Additionally, the blue symbols indicate results obtained at ambient pressure from References 35 and 38. The solid (and dashed) lines correspond to the outcomes of our DFT calculations for the zircon (and scheelite) phases. The associated errors are less than the dimensions of the symbols.

3.3 *Mechanical stability*

Our analysis confirms the mechanical stability of the zircon and scheelite structures. The computed elastic constants for the two phases of ErVO$_4$ are detailed in **Tables 2** and **3**. The positive eigenvalues observed in both elastic tensors indicate the elastic stability of each phase. Additionally, the calculated constants meet the Born criteria [43], further affirming the mechanical stability of these structures. For the zircon-type phase, our findings are consistent with earlier results obtained from Brillouin scattering measurements conducted on single crystals [44]; refer to Table 2. An interesting observation is that in the zircon structure $C_{33} > C_{11}$ and conversely in the scheelite structure $C_{11} > C_{33}$. This observation is consistent with the anisotropy of the axial compressibility we discussed in the previous section.

The elastic moduli of the two phases of ErVO$_4$ have been derived from the elastic constants by averaging the results obtained through the Hill approximation [45]. The computed bulk moduli for both phases are consistent with the values obtained from the prior equation of state analysis. In both phases, the Young's modulus and the bulk modulus differ by only 5%, suggesting that zircon and scheelite display comparable tensile and compressive stiffness along their length relative to their resistance to bulk compression. Additionally, it has been noted that the shear modulus in both phases is considerably lower than the bulk modulus, indicating a tendency towards shear deformations rather than volume contraction. This characteristic makes both phases of ErVO$_4$ susceptible to significant non-hydrostatic stresses [46,47]. This observation is consistent with the discussion we made on the fact we observed the stability of scheelite at higher pressures than in previous experiments. Additionally, the B/G ratio, which exceeds 1.75, allows us to infer that the two phases of ErVO$_4$ possess ductile characteristics. The values of the Poisson's ratio (ν) further support this assertion [48]. Finally, the Debye temperature ($\Theta_D$) for both phases were also calculated. The obtained temperatures are given in **Tables 2** and **3**. They are higher than 420 K indicating that the temperatures of interest for this study the properties if the crystals can be described by a quasi-harmonic approximation.

**Table 2**: Calculated elastic constants $C_{ij}$ (in GPa) for zircon-type ErVO$_4$. Additionally, the values for the bulk modulus (B), shear modulus (G), Young's modulus (E), Poisson's ratio (ν), the B/G ratio, and the Debye temperature ($\Theta_D$) are also presented. The elastic constants are compared with results from previous Brillouin scattering experiments [46].

| $C_{ij}$ | This work | Experiments [46] | Property | This work |
|---|---|---|---|---|
| $C_{11}$ | 250.96 | 256.6±5.11 | B | 139.25 |
| $C_{12}$ | 50.32 | 53±3 | G | 53.52 |
| $C_{13}$ | 87.90 | 79±6 | E | 142.02 |
| $C_{33}$ | 319.06 | 313±6 | ν | 0.330 |
| $C_{44}$ | 46.45 | 50.1±1.0 | B/G | 2.604 |
| $C_{66}$ | 20.66 | 17.7±0.9 | $\Theta_D$ | 423.75 |

**Table 3**: Calculated elastic constants $C_{ij}$ (in GPa) for zircon-type ErVO$_4$. Additionally, the values for the bulk modulus (B), shear modulus (G), Young's modulus (E), Poisson's ratio (ν), the B/G ratio, and the Debye temperature ($\Theta_D$) are also presented.

| $C_{ij}$ | This work | Property | This work |
|---|---|---|---|
| $C_{11}$ | 259.70 | B | 156.74 |
| $C_{12}$ | 128.69 | G | 61.30 |
| $C_{13}$ | 114.60 | E | 162.66 |
| $C_{33}$ | 224.07 | ν | 0.327 |
| $C_{44}$ | 58.13 | B/G | 2.557 |
| $C_{66}$ | 68.44 | $\Theta_D$ | 444.00 |

## 5. Conclusions

We have investigated the effect of pressure on the crystal structure of ErVO$_4$ through the use of single-crystal X-ray diffraction techniques. Experiments were performed at the ESRF combining synchrotron radiation and the use of the diamond-anvil cell. The experiments were carried out under quasi-hydrostatic conditions using helium as pressure medium. We found the zircon-scheelite transition at 7.9(1) GPa. In contrast with previous powder X-ray diffraction studies no phase coexistence was observed in the present experiments. We have also excluded the possibility of an intermediate phase existing between zircon and scheelite. We also determined the compressibility of the crystallographic axes of both structures and a room-temperature isothermal equation of state for zircon and scheelite. We have discussed our results in comparison with previous studies and with density-functional theory calculations we performed.

## DATA AVAILABILITY

The data that support the findings of this study are available from the corresponding author upon reasonable request.

## AUTHOR INFORMATION

**Corresponding Author**

*daniel.errandonea@uv.es

**ORCID**

Josu Sánchez-Martín: 0000-0003-0241-0217

Gastón Garbarino: 0000-0003-4780-9520

Samuel Gallego-Parra: 0000-0001-6516-4303

Alfonso Muñoz: 0000-0003-3347-6518

Sushree Sarita Sahoo: 0009-0000-0604-0765

Kanchana Venkatakrishnan: 0000-0003-1575-9936

Ganapathy Vaitheeswaran: 0000-0002-2320-7667

Daniel Errandonea: 0000-0003-0189-4221

**Author Contributions**

J.S.-M.: Experimental measurements, Investigation, Formal Analysis, Writing – review & editing. G.G. and S.G.-P.: Experimental measurements, Investigation, Writing – review & editing. A.M, S.S.S., K.V. and G.V.: Investigation, Writing – review & editing. D.E.: Conceptualization, Funding acquisition, Investigation, Project administration, Writing – original draft, Writing – review & editing. The manuscript was written through contributions of all authors. All authors have given approval to the final version of the manuscript.

**Notes**

The authors declare no competing financial interest.

**ACKNOWLEDGMENTS**

The authors thank Marco Bettinelli from Univ. of Verano for growing the single crystal of ErVO$_4$ used for this study. They also acknowledge the financial support from the Spanish Ministerio de Ciencia, Innovación y Universidades, MCIU, (DOI: 10.13039/501100011033) under Projects PID2022-138076NB-C41/44, and RED2022-134388-T and from Generalitat Valenciana through grants PROMETEO CIPROM/2021/075 and MFA/2022/007. J.S.-M. acknowledges the Spanish MCIU for the PRE2020-092198 fellowship. This study forms part of the Advanced Materials program and is supported by MCIU with funding from European Union Next Generation EU (PRTR-C17.I1) and by Generalitat Valenciana. The authors thank the ESRF for providing beamtime for the experiments. S.S.S. and K.V. would like to acknowledge IIT Hyderabad for its computational facility and the CSIR project with sanction No. 03(1433)/18/EMR-II for financial support. K.V. would like to acknowledge DST-FIST (SR/FST/PSI-215/2016) for financial support. G.V. would like to thank CMSD, University of Hyderabad for providing the computational facility and the Institute of Eminence, University of Hyderabad (UOH-IOE-RC3-21-046), for their financial support.
**REFERENCES**

[1] Far, B. F.; Maleki-Baladi, R.; Fathi-Karkan, S.; Babaei, M.; Sargazi, S. Biomedical applications of cerium vanadate nanoparticles: a review. *J. Mater. Chem. B* **2024**, *12*(3), 609-636. DOI: 10.1039/D3TB01786A

[2] Yang, R.; Chen, L.; Li, B. A new rare-earth orthovanadate microwave dielectric ceramic: ErVO$_4$. *Mat. Chem. Phys.* **2023**, *301*, 127630. DOI: 10.1016/j.matchemphys.2023.127630

[3] Che, Y.; Sun, S.; Tang, P.; Chen, H.; Ding, Y. Preparation of EuVO$_4$ by Hydrothermal Method and Its Photocatalytic Activity. *Integr. Ferroelectr.* **2022**, *227*(1), 231 - 234. DOI: 10.1080/10584587.2022.2065589

[4] Ruskuc, A.; Wu, C. J.; Rochman, J.; Choi, J.; Faraon, A. Nuclear spin-wave quantum register for a solid-state qubit. *Nature* **2022**, *602*, 408 - 413. DOI: 10.1038/s41586-021-04293-6

[5] Errandonea, D. High pressure crystal structures of orthovanadates and their properties. *J. Appl. Phys.* **2020,** *128*(4)*,* 040903. DOI: 10.1063/5.0016323
22